\documentclass[epj]{svjour}
\usepackage{graphics}
\usepackage{graphics}
\usepackage{graphicx}

\newcommand{\be}{\begin{eqnarray}}
\newcommand{\ee}{\end{eqnarray}}

\begin{document}
\title{Potential Models for Quarkonia}
\subtitle{What can we learn from potential models at finite temperature?}
\author{{\'Agnes M\'ocsy\thanks{\email{amocsy@pratt.edu}}}}
\institute{ Pratt Institute, Department of Math and Science, Brooklyn, NY 11205 USA, {\it and} \\
RIKEN Brookhaven Research Center, Brookhaven National
  Laboratory, Upton, NY 11973 USA}
\date{Received: date / Revised version: date}
%
\abstract{
In this paper I discuss what we can learn about quarkonium dissociation from lattice-potential based models. Special emphasis is given to results obtained in agreement by different models, and to the relevance of lattice QCD for potential models. Future directions are also discussed. %
\PACS{
      {PACS-key}{discribing text of that key}   \and
      {PACS-key}{discribing text of that key}
     } 
} 
\maketitle
%
\section*{Why are we interested? }
\label{intro}

One of the aims of relativistic heavy ion collisions is to produce quark-gluon plasma (QGP), a state of matter in which the constituents of our hadronic world are deconfined. Deconfinement is expected to happen at large energy densities, which can be obtained by heating matter to extreme high temperatures. High energy density matter has been already produced at SPS CERN and at RHIC BNL, and will be produced at the LHC, which just started its operation at CERN. To have control over what temperatures are achieved and whether deconfined matter has been produced we need a thermometer. The sequential melting of quarkonium has been long considered to be exactly that: the QGP thermometer \cite{satz}.  In a deconfined matter the force between the constituents of a quarkonium state, a heavy quark and its antiquark, is weakened by the color screening produced by the light quarks and gluons. For twenty years it has been believed that this screening leads to the dissociation (melting) of quarkonium \cite{matsui}. The different quarkonium states are expected to melt sequentially, at different temperatures. A suppressed yield of quarkonium can be visible in the dilepton spectrum, which is measured in experiments. 
 
$J/\psi$  suppression has been indeed measured by the different experiments \cite{exp}.  Understanding the data, however, turned out to be more complicated. The reason is that  the suppression pattern seen is not only due to the hot medium effects of screening, but more like due to the interplay of this with effects of cold nuclear matter \cite{cold}, as well as those of recombination \cite{reco}. In order to disentangle these different effects we must  know the properties of quarkonium in-medium and determine their dissociation temperatures.  

In principle, everything about a given quarkonium channel is embedded in its spectral function: The position of a peak in the spectral function corresponds to the mass of a bound state, while its width determines its lifetime. Melting of a state corresponds to the disappearance of a peak. A spectral function also contains information about the continuum and its threshold. So following how the spectral function changes with temperature can give us a theoretical insight to the temperature-dependence of quarkonium properties. There are two main lines of theoretical studies to determine quarkonium spectral functions at finite temperature: potential models and lattice QCD. Potential models have been widely used to study quarkonium, but their applicability at finite temperature is still under scrutiny. Lattice QCD provides the most  straightforward way to determine spectral functions, but the results suffer from discretization effects and statistical errors, and thus are still inconclusive. 

In the rest of the paper I discuss our current understanding of what we can learn using the potential models at finite temperature currently on the market, highlighting the agreement produced by the different groups, as well as the usage of lattice data (as input or as constrain) in context of these models.  

\section*{What Goes Into Potential Models}

Potential models are based on the assumption that the interaction between the heavy quark and its antiquark inside the quarkonium can be described by a potential. Due to the largeness of the heavy quark mass, $m_{c,b}\gg\Lambda_{QCD}$  and the smallness of the heavy quark velocity, $v\ll 1$, one can treat the quark-antiquark system nonrelativistically and solve the Scr\"odinger equation to obtain the bound state  properties.  

The zero temperature potential model with the Cornell potential has experienced great success: It describes well the experimentally observed quarkonium spectroscopy \cite{eichten}; It is verified on the lattice \cite{necco}; and it can be derived directly from QCD \cite{pnrqcd}. The latest is possible due to the hierarchy of well separated energy-scales $m\gg mv\gg mv^2$, which allows to systematically integrate out the different scales and obtain the non-relativistic potential QCD (pNRQCD) (in this the Cornell potential shows up as the zeroth order matching coefficient) \cite{pnrqcd}. 

Inspired by its success at zero temperature the potential model has been applied at finite temperature, with the main assumption that medium effects can be accounted for as a temperature-dependent potential. A few years ago it has become clear that instead of just looking at the individual bound states (procedure good at $T=0$ where quarkonium is well defined), we should rather obtain a unified treatment of bound states, threshold and continuum by determining the spectral function \cite{mocsyold}. Spectral functions can be obtained in a number of ways: by determining all the discrete states from solving the Schr\"odinger equation \cite{alberico}; or by using a T-matrix approach \cite{rapp}; or, and this has been our choice, by solving the Schr\"odinger equation for the nonrelativistic Green's function, and then make use of the optical theorem \cite{mocsyPRD}. In all these approaches one has to correctly account for the relativistic part of the continuum \cite{mocsyPRD}.  

At first, potential models at finite temperature used a phenomenological screened Cornell potential \cite{KMS}. This is all right for merely qualitative purposes. In order to get quantitative about how the bound state properties (e.g. binding energy) change with temperature, knowledge of the exact potential at finite $T$ is required. In principle, one should derive this directly from QCD, just as this has been done at zero temperature. Such derivation, however, is complicated by the existence of temperature-driven scales, $T, gT, g^2T~$, and has only been addressed recently \cite{vairo}. 
So in lack of knowledge resuming to different phenomenological potentials came naturally. In particular, it became popular using so-called lattice-based potentials, i.e. potentials constrained by lattice data on the free energy of a static quark-antiquark pair.  More precisely, by the change in the free energy of a medium at a given temperature when a static quark-antiquark pair is immersed into it. The lattice results show that above deconfinement the range of interaction between the quark and antiquark is strongly reduced, and this effect can be well described by exponential screening \cite{kaczmarek}. 

\begin{figure}[htbp]
\resizebox{0.48\textwidth}{!}{%
\includegraphics{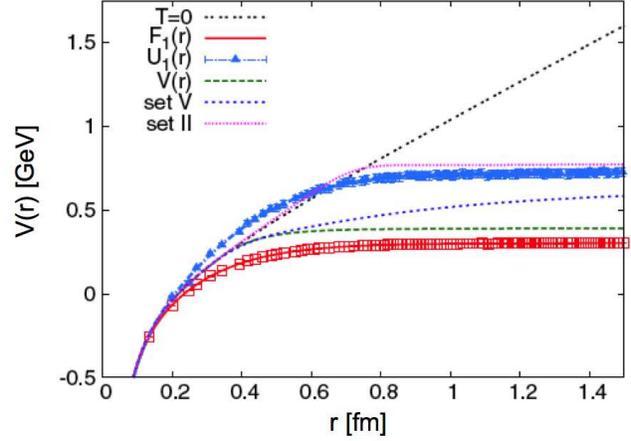}
}
\caption{Set of potentials allowed by lattice data on free energies. (See \cite{mocsyold} for details)}
\label{fig:potentials}
\end{figure}

One of the most debated questions of recent years is which lattice-based potential is to be used in the Schr\"odinger  equation. First, it was the free energy $F_1$ \cite{digal}. Then, it has been understood that this serves merely as a lower limit, since it contains an entropy contribution \cite{kaczmarek}. Removing the entropy the internal energy $U_1$ is obtained \cite{kaczmarek}. $U_1$ has been then used as potential \cite{rapp}. The interpretation of $U_1$ as potential is also questionable\footnote{$U_1$ includes medium polarization effects; has a huge increase near $T_c$; and has an unmotivated increased strength at short distances compared to the $T=0$ potential \cite{kaczmarek}.}, thus serving only as a sort of upper limit. Other lattice-based potentials on the market include the one proposed by Wong as a combination of $F_1$ and $U_1$ \cite{wong,alberico}, and a set of potentials constructed using the general features of the lattice free energy:  at short distances no deviation form the vacuum potential, at large distances exponential screening \cite{mocsyPRD}. A set of lattice-based potentials is shown in the left panel of figure \ref{fig:potentials}. 
  
\section*{What Comes Out of Potential Models}

It is clear that different potentials can lead to different spectral functions and thus different properties for the quarkonium states. More precisely, the deeper the potential is the stronger the binding of quarkonium can be, and the higher the possible dissociation temperatures might then be. 

\begin{figure}[htbp]
\resizebox{0.5\textwidth}{!}{%
\includegraphics{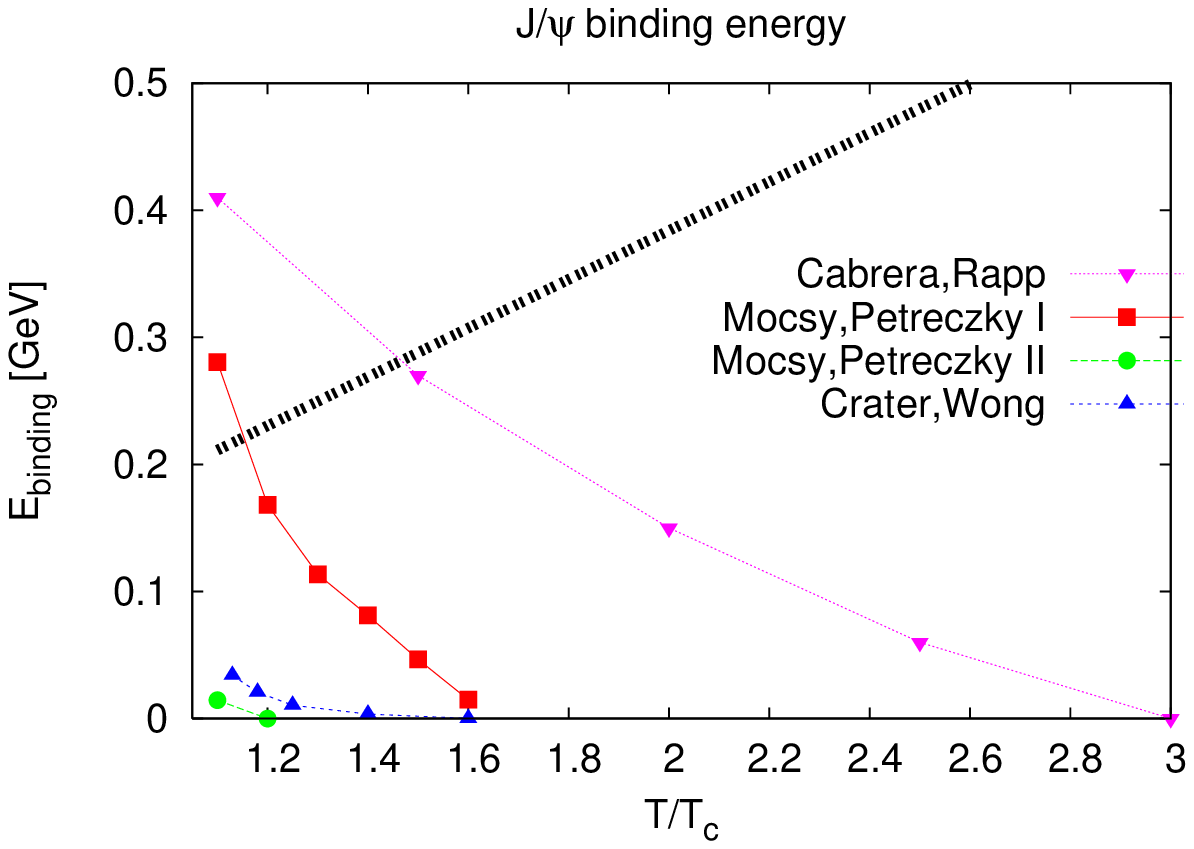}
} \\
\resizebox{0.5\textwidth}{!}{%
\includegraphics{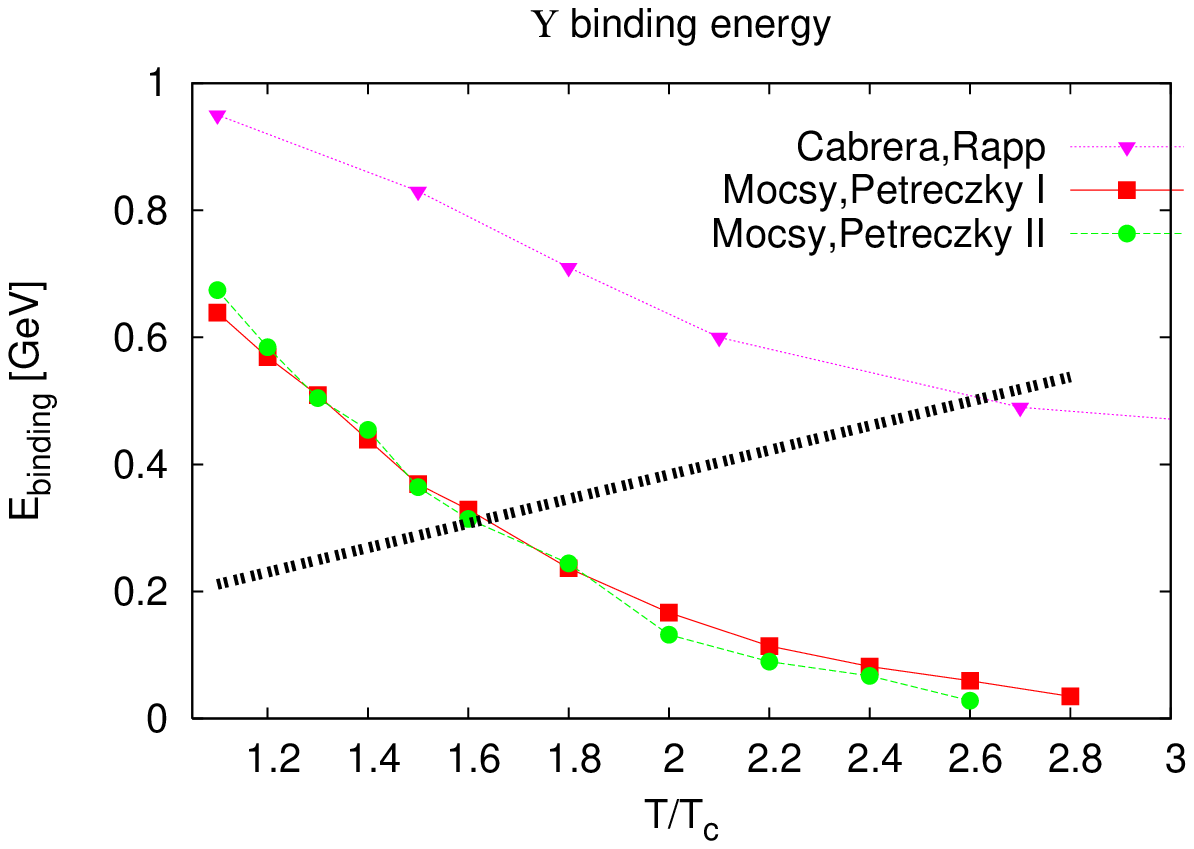}
} 
\caption{The binding energy of the $J/\psi$ (upper panel) and the $\Upsilon$ (lower panel) determined in different potential models. The black line corresponds to the temperature line. }
\label{fig:binding}
\end{figure}

Granted the differences, there are, however, a number of essential features common for all the spectral functions from potential models \cite{alberico,rapp,mocsyPRD}: \\
1) There is a large threshold (rescattering) enhancement beyond what corresponds to free quark propagtion. This enhancement is present even at high temperatures, and it is the  
indication that there is correlation persisting between the quark and antiquark. Threshold enhancement has been identified in all of the channels (charmonium and bottomonium S- and P-states). \\
 2) There is a strong decrease with increasing temperature of the binding energies (the distance between peak position and continuum threshold in the spectral function) determined from potential models. This is illustrated in Figure \ref{fig:binding} for the  $J/\psi$ (upper panel) and the $\Upsilon$ (lower panel). The black line on the figure corresponds to the temperature line, where the binding energy becomes of the order of temperature. 

What we learn from this is that one can obtain a spectral function that exhibits a resonance-like peak, but the corresponding binding energy can be small. It becomes meaningless, for instance, to talk about a $3~$GeV mass $J/\psi$ with $20~$MeV binding energy! Therefore, unlike for quarkonium at zero temperature, at high temperature it is incorrect to call a state dissociated only when it's binding energy becomes zero. The condition for a state to be considered dissociated must be thus weaker than $E_{bin}=0~$. We discuss this in more detail later in this paper. At this point let me emphasize, that with increasing temperature a state can become quickly broadened (weakly bound) and thermal fluctuations can dissociate it. Thermal broadening of quarkonium, although not taken into account in potential model spectral functions, has been addressed in a number of independent calculations: NLO perturbative QCD \cite{park}, QCD sum rule \cite{morita}, and resummed perturbative QCD \cite{laine}. All of these calculations show that, for example, the $J/\psi$ is significantly broadened at temperatures right above that of deconfinement. 

It is important to keep in mind when looking at spectral functions from potential models using lattice-based potentials: none of these calculations include the true width of a state. Therefore just seeing the peak structure in these spectral functions is incomplete on its own and can be misleading. Note however, that in the perturbative approach the potential has also an imaginary part \cite{vairo,laine,blaizot}. And the inclusion of the imaginary part clearly broadens the peak \cite{laine}.  
  
Besides illustrating the decrease of binding energies with increasing temperature, Figure \ref{fig:binding} also shows that there are large uncertainties from the modeling of the potential. In the following I will discuss whether we can distinguish which potential is the "right potential", if any, by making further use of lattice data.

\section*{Potential Models $\&$ Lattice QCD} 

We have discussed that in lack of knowing the finite temperature quark-antiquark potential in QCD (pNRQCD) the free energy of a static quark-antiquark pair calculated on the lattice inspired a series of lattice-based potentials used as input for potential models. 
\begin{figure*}[]
\begin{center}
\resizebox{0.65\textwidth}{!}{%
\includegraphics{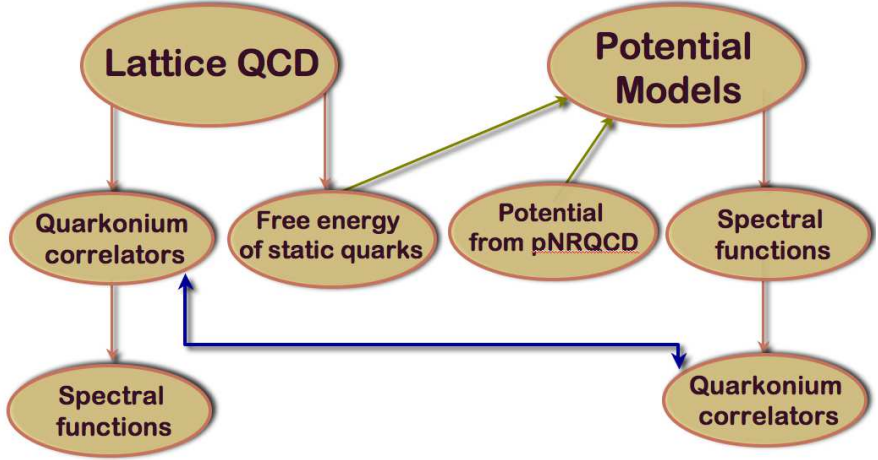}
}%
\end{center}
\caption[]{Structural chart of lattice QCD and potential model calculations.}
\label{fig:chart}
\end{figure*} 
There are, however, two independent lattice QCD calculations relevant for quarkonium studies: One is the calculation of the T-dependence of the free energy of a static quark-antiquark pair \cite{kaczmarek} , and the other is the calculation of the current-current correlation function of mesonic currents in Euclidean-time. From the current-current correlation functions for different quarkonium channels the quarkonium spectral functions are {\it extracted} \cite{lattice}. The extraction of spectral function using the Maximum Entropy Method is still difficult due to discretization effects and statistical errors. So the uncertainties in the spectral function are significant and details of this cannot be resolved! Thus Òit is difficult to make any conclusive statement based on the shape of the spectral functionsÓ \cite{jako}. 

Contrary to the extracted spectral functions, the lattice correlators are determined reliably. Therefore, it has been suggested to compare correlators from potential models to the ones from lattice QCD \cite{mocsyold}. Correlators in potential models can be calculated using the integral representation of these in terms of the spectral function, 
\be
G(\tau,T)=\int d\omega\sigma(\omega,T)\cosh\left(\omega(\tau-1/2T)\right)/\sinh\left(\omega/2T\right)\, . \nonumber
\ee 
It is customary to present the temperature-dependence of the correlators as the ratio $G/G_{recon}~$, where $G_{recon}(\tau,T=0)$ is the correlator ÒreconstructedÓ from spectral function at zero (or some low) temperature. This ratio evaluated in potential models should then be compared to this ratio directly calculated on the lattice. 

The chart in Figure \ref{fig:chart} shows a diagram of the interconnectedness of lattice QCD and potential models. Note though, that these approaches describe quarkonium spectral functions in equilibrated QGP. In order to bridge these theoretical approaches to experimental data one needs to resume to some dynamical model (see for instance \cite{shuryak}).

\begin{figure}[htbp]
\resizebox{0.48\textwidth}{!}{%
\includegraphics{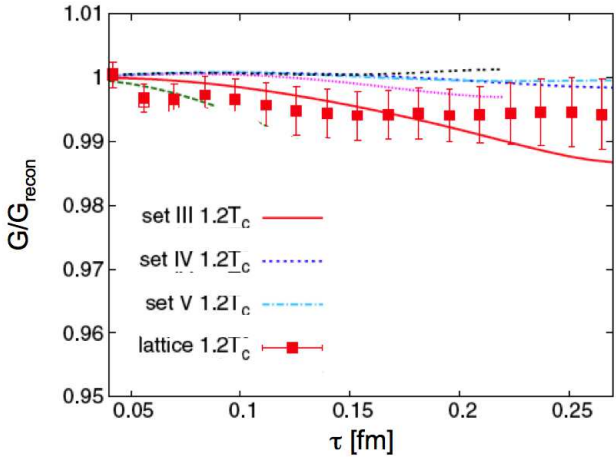}
} 
\caption{The correlator ratio obtained with the set of potentials from Fig. \ref{fig:potentials} compared to lattice data (right) at $1.2T_c~$. (See \cite{mocsyold} for details)}
\label{fig:result}
\end{figure}

So the idea is to constrain different potential models using data from lattice \cite{mocsyPRD}. 
The approach is the following: Assume a lattice-based potential and determine the corresponding spectral function in a given quarkonium channel. From the spectral function determine the ratio of correlators $G/G_{rec}$ and compare it to lattice data. The "correct" potential and spectral function would be the one in best agreement with the data. 
The surprising result is shown in Figure \ref{fig:result}. This figure displays the correlator ratios in the pseudoscalar charmonium channel at $1.2T_c$ obtained using the set of potentials within the allowed ranges shown in Figure \ref{fig:potentials} (except for $U_1$), and compared to lattice data from \cite{jako}. 
The surprise is that the complete set of potentials all provide agreement of $1-2 \%$ accuracy with lattice correlators, yielding indistinguishable results. Using the internal energy $U_1$ as potential yields worst results \cite{rapp}. 
Interestingly, spectral functions that exhibit a ground state peak up to higher temperatures, obtained with the more binding potentials, and spectral functions with no resonance-like structure seen already near $T_c$, obtained with less binding potentials,  (see spectral functions in \cite{mocsyPRD}) all yield flat correlator ratios. 
This is understood as the threshold enhancement compensates for the melting of states keeping the correlators flat. Thus possible dramatic changes in the spectral function are not reflected in the correlator. We therefore cannot identify the "correct" potential, cannot determine the exact spectral function and quarkonium properties from such comparisons.  

The agreement to the correlator ratio is also good for bottomonium S-wave, as well as for all the P-states \cite{mocsyPRD}. It is now understood that the dominant source of the T-dependence of the correlator ratios comes from the commonly overlooked zero-modes \cite{umeda}. These are low-frequency contributions to the spectral functions at finite T describing the scattering states of single heavy quarks, in addition to the usual bound and unbound quark-antiquark pairs in a given channel.  The zero-mode contribution is understood in terms of quasi-free quarks with some effective mass \cite{peterQM}, indicating the presence of free heavy quarks in the deconfined phase. Furthermore, the zero-mode gives a constant contribution to correlator. One can eliminate it by looking at ratio of the derivatives of correlators \cite{umeda}.  The high energy part which carries info about bound states is flat, i.e. shows almost no T-dependence until $3T_c$ in all of the channels! One can conclude that the flatness is not related to survival, since it would also imply that the $\chi_c$ survives until $3T_c$. The understanding becomes then simple: the dramatic changes seen in the spectral function are not reflected in the correlator. In case of the $\Upsilon$ this means that the spectral function does not change to higher temperatures, the state survives, and the correlator ratio stays flat. For all the other channels ($J/\psi$, $\chi_c$, $\chi_b$, etc) the spectral enhancement near threshold compensates for the melting of states keeping the correlators flat. Let me make this statement more transparent through the discussion in the next section. 

\section*{So What Have We Learned?} 

As discussed earlier, it has been customary to consider a state dissociated when its binding energy becomes zero. In principle, a state is dissociated when no peak structure is seen, but the widths shown in spectral functions from current potential model calculations are not physical. Broadening of states as the temperature increases is not included in any of these models. At which T the peak structure disappears then? In \cite{mocsyPRL} we argue that no need to reach $E_{bin}=0$ to dissociate, but when $E_{bin}< T$ a state is weakly bound and thermal fluctuations can destroy it. Let us quantify this statement. 

\begin{figure}{}
\begin{center}
\parbox{40mm}{
\vspace{0mm}
  \includegraphics[width=0.28\textwidth]{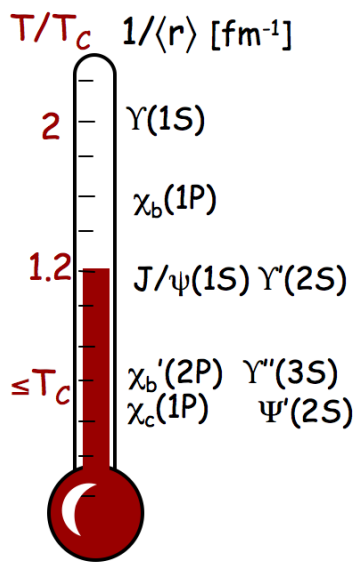}
\parbox{0.5\textwidth}{
\vspace{0mm}\caption[]{ The QGP thermometer. \hspace{0.99\textwidth}}\label{fig:thermometer}}
}
\end{center}
\end{figure}

Due to the uncertainty in the potential we cannot determine the binding energy exactly, but we can nevertheless set an upper limit for it  \cite{mocsyPRL}: We can determine $E_{bin}$ with the most confining potential that is still within the allowed ranges by lattice data on free energies. For the most confining potential the distance where deviation from $T=0$ potential starts is pushed to large distances so it coincides with the distance where screening sets in \cite{mocsyPRD}. From $E_{bin}$ we can then estimate, following \cite{dima}, the quarkonium dissociation rate due to thermal activation, obtaining this way the thermal width of a state $\Gamma(T)$.  
At temperatures where the width, that is the inverse of the decay time, is greater than the binding energy, that is the inverse of the binding time, the state will likely to be dissociated. In other words, a state would melt before it binds. For example, already close to $T_c$ the $J/\psi$ would melt before it would have time to bind. To quantify the dissociation condition we have set a more conservative condition for dissociation: $2E_{bin}(T)<\Gamma(T)$. The result for different charmonium and bottomonium states is shown in the thermometer of figure \ref{fig:thermometer}. Note, that all these numbers are to be though of as upper limits.  

In summary, potential models utilizing a set of potentials between the lower and upper limit constrained by lattice free energy lattice data yield agreement with lattice data on correlators in all quarkonium channels. Due to this indistinguishability of potentials by the data the precise quarkonium properties cannot be determined this way, but the  upper limit can be estimated. The decrease in binding energies with increasing temperature, observed in all the potential models on the market, can yield significant broadening, not accounted for in the currently shown spectral functions from these models.  The upper limit estimated using the confining potential predicts that all bound states melt by $1.3T_c$, except the Upsilon, which survives until $2T_c$.  The large threshold enhancement above free propagation seen in the spectral functions even at high temperatures, again observed in all the potential models on the market, compensates for melting of states (yielding flat correlators), and indicates that correlation between quark and antiquark persists.  Lattice results are thus consistent with quarkonium melting.

\section*{And What's Next?}

Implications of the QGP thermometer of figure \ref{fig:thermometer} for heavy ion collisions should be considered by phenomenological studies. This can have consequences for the understanding of the $R_{AA}$measurements, since now the $J\psi$ should melt at SPS and RHIC energies as well. The thermometer also suggests that the $\Upsilon$ will be  suppressed at the LHC, and that centrality dependence of this can reveal whether this happens already at RHIC. So measurements of the $\Upsilon$ can be an interesting probe of matter at RHIC as well as at the LHC. 

The exact determination of quarkonium properties the future is in the effective field theories from QCD at finite T. First works on this already appeared \cite{vairo} and both real and imaginary parts  of the potential have been derived in certain limits. In these works there is indication that most likely charmonium states dissolve in QGP due thermal effects, such as activation to octet states, screening, Landau-damping. 

The correlations of heavy-quark pairs that is embedded in the threshold enhancement should be taken seriously and its consequences, such as possible non-statistical recombination taken into account in dynamic models that attempt the interpretation of experimental data \cite{shuryak}.  

All of the above discussion is for an isotropic medium. Recently, the effect of anisotropic plasma has been considered \cite{adrian}. Accordingly, quarkonium might be stronger bound in an anisotropic medium, especially if it is aligned along the anisotropy of the medium (beam direction). Qualitative consequences of these are considered in an upcoming publication \cite{anisotrop}. Also, all of the above discussion refers to quarkonium at rest. Finite momentum calculations are under investigation. It is expected that a moving quarkonium dissociates faster.

\section*{Acknowledgment}

I thank P\'eter Petreczky for collaboration on performing this work and for carefully reading this manuscript.  I am grateful to the Organizers of Hard Probes 2008 for inviting me to give this talk. This work has been supported by U.S. Department of Energy under Contract No. DE-AC02-98CH10886.


\vfill\eject
\end{document}